\documentclass[twocolumn,superscriptaddress,amsmath,amssymb,aps,prl]{revtex4-1}

\usepackage{graphicx}
\usepackage{dcolumn}
\usepackage{bm}

\usepackage[colorlinks,urlcolor=blue,citecolor=blue,linkcolor=blue]{hyperref}

\newcommand{\up}{\uparrow}
\newcommand{\down}{\downarrow}
\renewcommand{\k}{{\bf k}}
\newcommand{\p}{{\bf p}}

\newcommand{\q}{{\bf q}}

\renewcommand{\u}{{\bf u}}
\renewcommand{\v}{{\bf v}}

\newcommand{\ket}[1]{\left|{#1}\right>}

\newcommand{\eb}{E_{\rm B}}
\newcommand{\eq}{\epsilon_{q}}

\newcommand{\beq}{\begin{equation}}
\newcommand{\eeq}{\end{equation}}

\newcommand{\sout}[1]{}

\begin{document}

\title{Impurity-induced multi-body resonances in a Bose gas}

\author{Zhe-Yu Shi}
\thanks{These two authors contributed equally to this work. The order was determined by a digital coin toss.}
\affiliation{School of Physics and Astronomy, Monash University, Victoria 3800, Australia}

\author{Shuhei M. Yoshida}
\thanks{These two authors contributed equally to this work. The order was determined by a digital coin toss.}
\affiliation{School of Physics and Astronomy, Monash University, Victoria 3800, Australia}
\affiliation{Department of Physics, The University of Tokyo, Tokyo 113-0033, Japan}

\author{Meera M.\ Parish}
\affiliation{School of Physics and Astronomy, Monash University, Victoria 3800, Australia}

\author{Jesper Levinsen}
\affiliation{School of Physics and Astronomy, Monash University, Victoria 3800, Australia}

\date{\today}

\begin{abstract}
We investigate the problem of $N$ identical bosons that are coupled to an impurity particle with infinite mass.
For non-interacting bosons, we show that a dynamical impurity-boson interaction, mediated by a closed-channel dimer, 
can induce an effective boson-boson repulsion which strongly modifies the bound states consisting of the impurity and $N$ bosons. In particular, we demonstrate the existence of two universal ``multi-body'' resonances, where all
multi-body bound states involving any $N$ emerge and disappear. 
The first multi-body resonance corresponds to infinite impurity-boson scattering length, $a\to +\infty$, while the second corresponds to the critical scattering length $a^*>0$ beyond which the trimer ($N=2$ bound state) ceases to exist. 
Crucially, we show that the existence of $a^*$ ensures that the ground-state energy in the multi-body bound-state region, $\infty>a> a^*$, is bounded from below, with a bound that is independent of $N$. 
Thus, even though the impurity can support multi-body bound states, they become increasingly fragile beyond the dimer state. This has implications for the nature of the Bose polaron currently being studied in cold-atom experiments.
\end{abstract}

\pacs{}

\maketitle

Recent advances in cold-atom experiments have enabled a variety of quantum impurity problems to be realized and investigated experimentally. Of particular interest is the case of an impurity particle interacting with a bosonic medium, 
since this scenario is central to our understanding of a wide range of systems, such as electrons coupled to phonons~\cite{mahan}, $^3$He--$^4$He mixtures~\cite{BaymPethick1991book}, and spins coupled to a dissipative environment~\cite{Leggett1987}. Already, there have been experiments addressing the canonical case of an impurity in a Bose-Einstein condensate (BEC) --- the so-called \emph{Bose polaron}~\cite{Hu2016,Jorgensen2016,Camargo2018}.  
However, there are currently conflicting theories about the behavior in the regime of strong boson-impurity attractive interactions.
One view is that the impurity binds many bosons to form a ``superpolaronic'' state~\cite{Tempere2009,Shchadilova2016,Schmidt2016b}, while other works consider the Bose polaron to be a highly correlated state that involves only a few particles in the dressing cloud~\cite{Rath2013,Li2014,Levinsen2015,Yoshida2018}.

In this Letter, we shed light on this issue by investigating
the problem of an infinitely heavy impurity interacting with $N$ identical bosons. 
In particular, we focus on impurity-boson interactions that are mediated by a closed-channel dimer state, as is the case for interatomic interactions  
tunable via a Fano-Feshbach resonance~\cite{Chin2010}. 
For non-interacting bosons, one might naively surmise that the ground state would be simply formed from bosons all occupying the single-particle ground state. 
However, while this would indeed be correct for a static scattering potential, this
picture neglects the possibility of boson-boson correlations induced by the dynamical nature of the boson-impurity interaction.
Specifically, once the impurity  
is in the closed-channel configuration, the interaction is unavailable to any other boson, 
resulting in an impurity-induced effective repulsion between the bosons. This constraint on the 
closed channel renders the problem non-trivial and precludes its exact solution for arbitrary boson number $N$.

\begin{figure}
	\includegraphics[width = 0.5\textwidth]
	{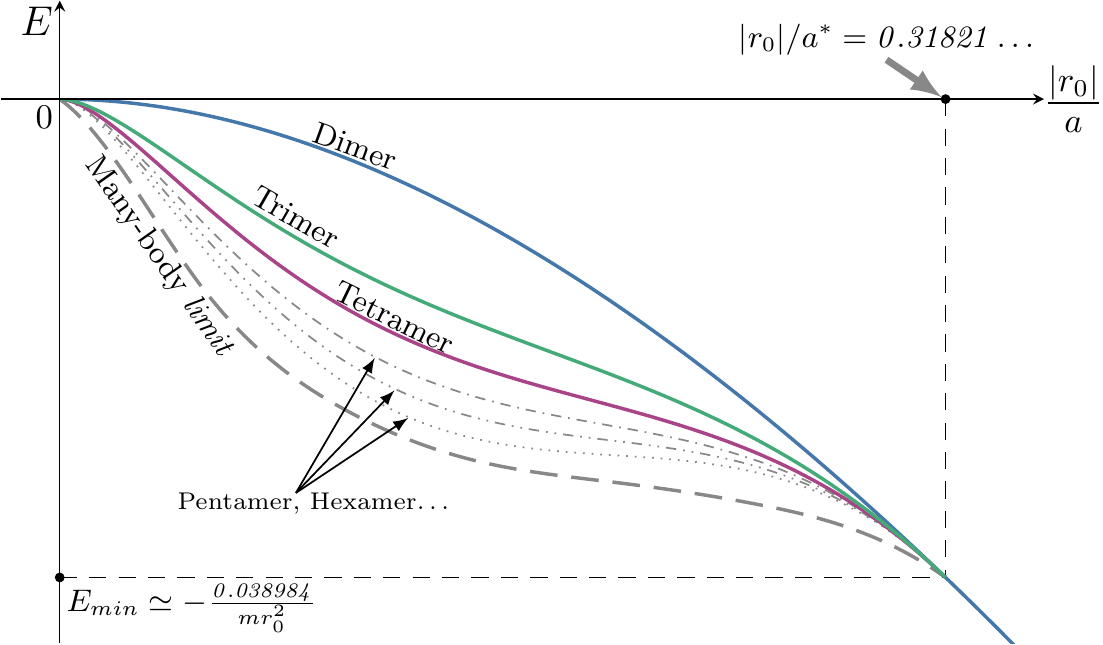}
	\caption{Schematic depiction of the bound-state spectrum for $N$ identical non-interacting bosons and one infinitely heavy impurity. Here $a$ and $r_0$ are the impurity-boson scattering length and effective range, respectively. The solid lines are based on analytical and exact calculations within our model (see text), while the dashed and dotted lines are conjectured multi-body bound states based on the asymptotic behavior at the multi-body resonances, 
	 $a = a^*$, $+\infty$. To enhance visibility, we have highly exaggerated the trimer and tetramer binding energies.}
	 \label{3fig}
\end{figure}

Motivated by the above,  we tackle this problem by examining the behavior of its bound states
with increasing $N$, as depicted in Fig.~\ref{3fig}.
Using a 
two-channel model for the impurity-boson interactions~\cite{Timmermans1999fri}, 
we can map the problem to an Anderson-like model, which allows us to solve the three-body
problem ($N=2$) \emph{analytically}. In particular, we obtain an analytic implicit
expression for the trimer energy at positive impurity-boson scattering lengths $a$, and we demonstrate that there is a
critical scattering length $a^*$ beyond which no trimer state exists. 
Indeed we find that both $a = a^*$ and $1/a=0$ correspond to ``multi-body'' resonances, which mark the emergence of multi-body bound states involving any $N$ (see Fig.~\ref{3fig}). 
A detailed analysis of the few-body problem is contained in Ref.~\cite{longpaper}.

We argue 
that these multi-body resonances are universal in the sense that they 
exist in any model with an effective three-body repulsion involving the impurity.
For the multi-body resonance at $a \to +\infty$, the 
energies take the universal
form $E_{N+1}/\eb=-N+\frac{N(N-1)\pi}{\log(a/r)}$, with $\eb>0$ the
two-body (dimer) binding energy, while the length scale $r$ under the logarithm depends
on the exact mechanism through which the effective boson repulsion is
generated. 
Furthermore, we find the same universal behavior at the second multi-body resonance $a = a^*$,
where the dimer state now plays the role of the impurity.
Crucially, we show that the existence of $a^*$
guarantees that the ground-state energy in the multi-body bound-state region is
bounded from below, with a bound that is independent of $N$, 
in stark contrast to the naive expectation that $E_{N+1}\sim N$.
We argue that this places strong constraints on the behavior of the Bose polaron in the strong-coupling regime.

\paragraph{Models.--}
To describe the scenario of a single fixed impurity atom interacting with bosons via a Feshbach resonance, we use
a two-channel Hamiltonian~\cite{Timmermans1999fri},
\begin{align}
H
&= \sum_\k \epsilon_\k b_\k^\dag b_\k 
    + \nu_0 d^\dag d
    + g \sum_\k \left(
        d^\dag c\, b_\k +d\, c^\dag b_\k^\dag
    \right).\label{2channel}
\end{align}
Here $b_\k^\dagger$ creates a bosonic atom with momentum $\k$ and mass $m$, while $c^\dagger$ ($d^\dagger$) creates an impurity atom (closed-channel dimer),   
and $\epsilon_\k=\frac{k^2}{2m} \equiv \frac{|\k|^2}{2m}$ is the boson dispersion 
(we set $\hbar$ and the system volume to 1).
Without loss of generality, we only take the localized modes of the impurity and closed-channel dimer at the origin, omitting the momentum or coordinate variable from $c$ and $d$.

According to Eq.~\eqref{2channel}, the impurity interacts with the bosons through forming a closed-channel dimer described by $d^\dagger$. 
To relate the coupling strength $g$ and the bare 
detuning $\nu_0$ to physical quantities, we solve the two-body problem and match the solution with the standard scattering amplitude $f_0(\k)=-(a^{-1}-r_0k^2/2+ik)^{-1}$, where $a$ is the $s$-wave scattering length and $r_0$ is the effective range. As a result, $\frac{m}{2\pi a}=-\frac{\nu_0}{g^2}+\sum_{\k}\frac{1}{\epsilon_\k}$ and $r_0=-\frac{2\pi}{m^2g^2}.$

For a single impurity, we require $c^\dag c+d^\dag d=1$.
Under this constraint, we may simplify the Hamiltonian \eqref{2channel} by making the replacement 
$d^\dagger c \to d^\dagger$ in the interaction term and then taking 
$d^\dag$ to be a fermionic operator~\footnote{It can be shown that $d^\dagger c$ and $d^\dagger$ have the same matrix elements in the subspace of $d^\dagger d+c^\dagger c=1$. Indeed, this may be viewed as an ``inverse'' slave boson transformation.}.
However, the resulting quadratic Hamiltonian is still non-trivial because the off-diagonal terms such as $b_\k d^\dag$ will mix bosonic and fermionic degrees of freedom. To resolve this issue, we 
instead take $d^\dag$ to be a bosonic operator and introduce an on-site repulsion to avoid its double occupancy,
\begin{align}
    H=&\sum_\k\epsilon_\k b_\k^\dagger b_\k+\nu_0 d^\dag d+g\sum_\k\left(d^\dag b_\k+d b_\k^\dag\right)\nonumber\\
    &+\frac{U}{2}d^\dag d^\dag dd,\hspace{2cm} U\rightarrow +\infty\label{Anderson}
\end{align}
where the first three terms are quadratic and hence can be easily diagonalized.
The induced repulsion in the last term only appears when two or more bosons occupy the closed channel state. However, as we demonstrate below, this seemingly harmless interaction makes the Hamiltonian completely nontrivial, and indeed there is no simple solution even for the three-body problem. In fact, 
Eq.~\eqref{Anderson} is closely related to the Anderson impurity model~\cite{anderson1961}, 
which has been studied intensively during the past few decades due to its relation to Kondo physics. Furthermore, Eq.~\eqref{Anderson} can be mapped to a spin-$1/2$ model via the transformation $d^\dag = (\sigma_x + i\sigma_y)/2$ and $d^\dag d = (\sigma_z + 1)/2$, where $\sigma_i$ are the Pauli spin matrices and the spin-$\down$ ($\up$) state denotes the absence (presence) of a closed-channel dimer. This then yields a variation of the so-called spin-boson model~\cite{Leggett1987}, which has been mapped to the Kondo model in certain regimes and which also cannot be solved exactly in general.

To gain insight into the impurity problem, we investigate few-boson systems 
and then infer the behavior in the limit $N \gg 1$.
To this end, 
we focus on the bound states supported by the impurity. Note that there is no Efimov effect~\cite{Efimov1970} when the impurity mass is infinite, and thus bound states only exist when the scattering length $a > 0$. 
In this case, 
the two-body spectrum consists of a continuum of scattering states as well as a bound state with binding energy $E_\text{B}\equiv\frac{\kappa_\text{B}^2}{2m}$, where $\kappa_\text{B}=(1-\sqrt{1-2r_0/a})/r_0$. The eigenstates are
\begin{align}
    B_\lambda^\dagger|0\rangle=\left(\sum_\p\zeta_{\lambda \p} b^\dagger_\p+\eta_\lambda d^\dagger\right) |0\rangle,\quad \lambda=\mathbf{k}\text{ or } i\kappa_\text{B},\label{def_B}
\end{align}
where $\lambda=\k$ corresponds to a scattering state with $E=\epsilon_\k$, while $\lambda=i\kappa_\text{B}$ corresponds to the bound state with energy $E_2=\epsilon_{i\kappa_{\text{B}}}=-E_{\rm B}$. We present the explicit form of the wave functions $\zeta_{\lambda \p}$ and $\eta_\lambda$ in Ref.~\cite{longpaper}.

The states $B_\lambda^\dagger|0\rangle$ form an orthonormal basis of the two-body Hilbert space, and thus $B_\lambda$ satisfies bosonic commutation relations. Therefore, we can
diagonalize the quadratic part of the Hamiltonian~\eqref{Anderson} and obtain 
\begin{align}
    H={\sum_{\k}}'\epsilon_\k B_\k^\dagger B_\k+\frac{U}{2} {\sum_{\substack{\k,\p\\ \u,\v}}}'
        \chi_{\k \p}^* \chi_{\u \v}
        B_{\k}^\dag B_{\p}^\dag B_{\u} B_{\v},\label{final_model}
\end{align}
where ${\sum_\k}'h(\k)$ is a short-hand notation for ${\sum_\k}'h(\k)\equiv\sum_\k h(\k)+h(i\kappa_\text{B})$, and 
$\chi_{\u \v} \equiv \eta_{\u}\eta_{\v}$.

The final Hamiltonian (\ref{final_model}) greatly simplifies our problem in two aspects. First, because we have eliminated the impurity, 
the $(N+1)$-body problem in the original model (\ref{2channel}) is effectively reduced to an $N$-body problem. Second, although we have introduced an extra interaction term, it has the form of a separable potential.
The separability of the interaction significantly simplifies the calculation of few-body properties 
and allows us to derive a series of analytical results~\cite{longpaper}, which we now discuss. 

\paragraph{Three-body problem.--} Within our model (\ref{final_model}), a general three-body state is 
$\ket{\psi}=\sum_{\k\p}'\varphi_{\k\p}B_\k^\dagger B_\p^\dagger|0\rangle$. Using the Schr\"{o}dinger equation we then obtain 
\begin{align}
    (E-\epsilon_\k-\epsilon_\p)\varphi_{\k\p} &=
    U{\sum_{\u,\v}}'\chi_{\k\p}^*\chi_{\u\v}\varphi_{\u\v}.\label{3-body eq.}
\end{align}
For a bound trimer, 
we assume $E<-\eb$ and define $f\equiv U\sum_{\u\v}'\chi_{\u\v}\varphi_{\u\v}$. As a consequence of the separability of the interaction, Eq.~(\ref{3-body eq.}) can be simplified into
\begin{align}
    \left[\frac{1}{U}-Z(E)\right]f=0,\quad Z(E)\equiv{\sum_{\k,\p}}'\frac{|\chi_{\k\p}|^2}{E-\epsilon_\k-\epsilon_\p}. \label{finiteU_eq}
\end{align}
For $E<-E_\text{B}$, the function $Z(E)$ can be calculated analytically 
and we present its explicit expression in Ref.~\cite{longpaper}. It is now safe to take the limit $U\rightarrow+\infty$ in Eq.~(\ref{finiteU_eq}), and the trimer energy $E_3$ is then simply given by the root of $Z(E_3)=0$. We solve this equation exactly~\cite{longpaper}, and find the existence of a single trimer for scattering lengths in the interval $a^*<a<\infty$, where the critical scattering length $a^*$ in Fig.~\ref{3fig} is related to the effective range by $|r_0|/a^*\simeq 0.31821\ldots$.

\begin{figure}
	\includegraphics[width = 0.45\textwidth]
	{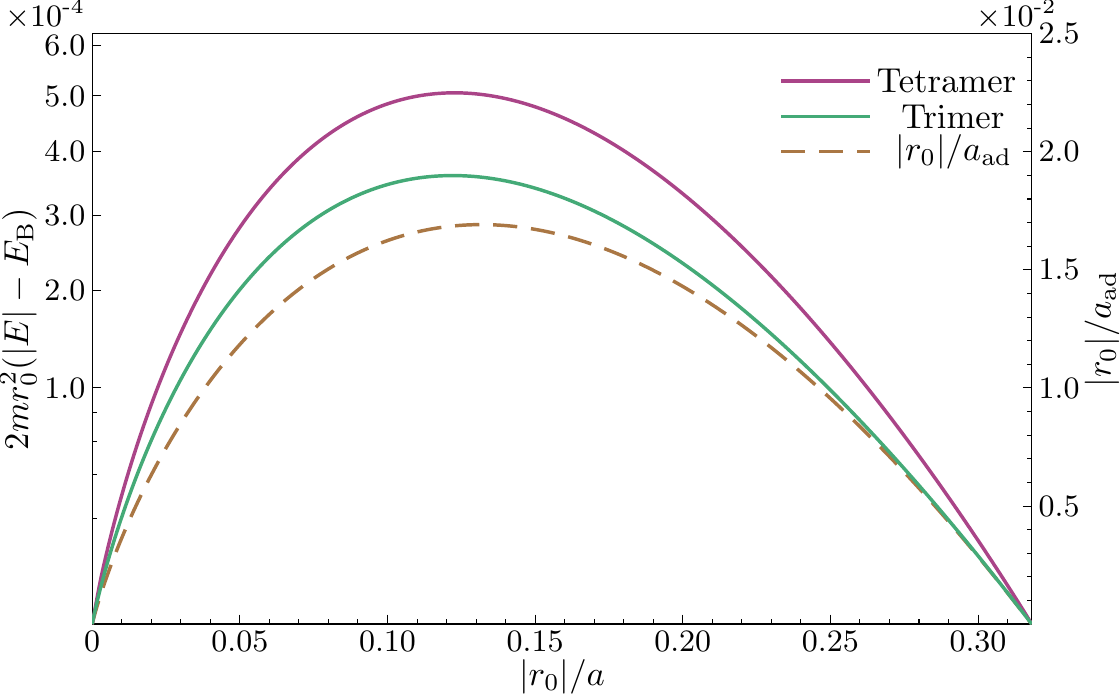}
	\caption{
	Trimer and tetramer energies (solid) for the two-channel model measured from the dimer energy as a function of $|r_0|/a$. We also show the inverse atom-dimer scattering length $|r_0|/a_{\text{ad}}$ (dashed) in the same region. We plot the energies in a quadratic scale while the inverse scattering length is shown in a linear scale, and thus the fact that $|r_0|/a_{\text{ad}}$ is very close to the trimer line indicates that the trimer binding energy is well approximated by $1/2ma_{\text{ad}}^2$.
	}\label{fig2}
\end{figure}

\begin{figure}
	\includegraphics[width = 0.45\textwidth]
	{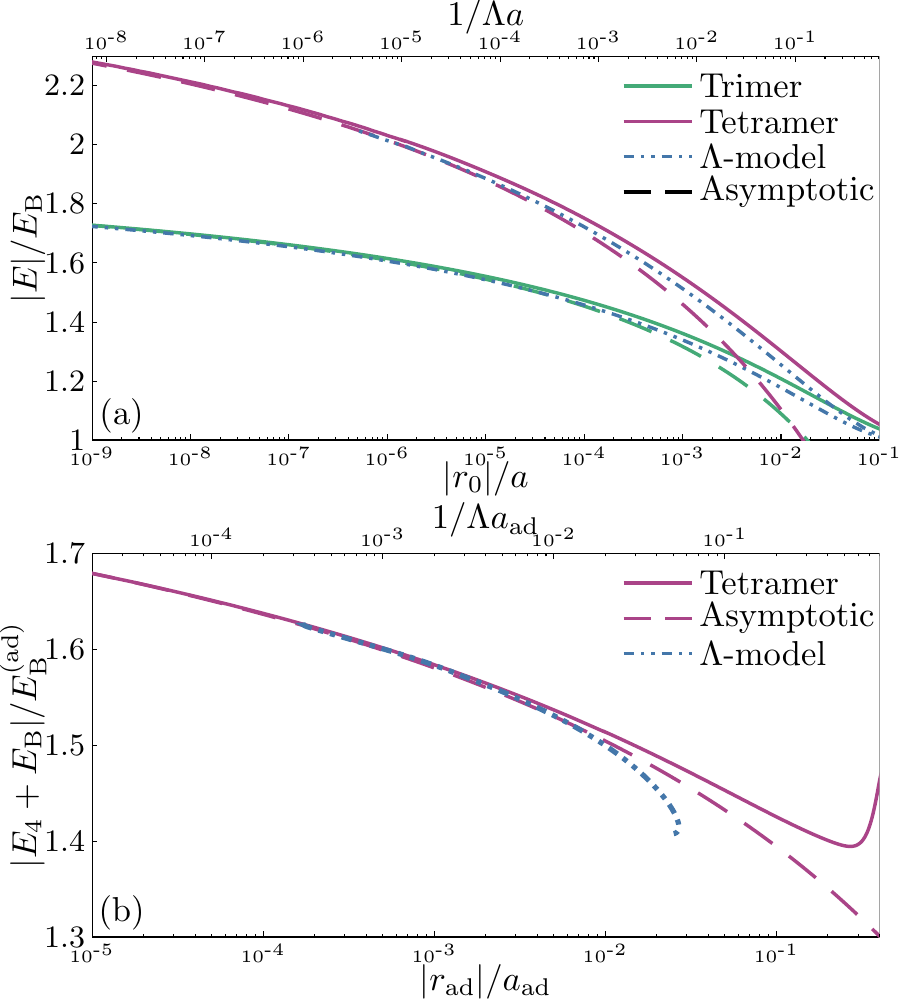}
	\caption{Trimer and tetramer energies 
	near $a=+\infty$ (a) and $a=a^*$ (b). We compare results from both the two-channel model (solid) and $\Lambda$-model (dash dotted), with the asymptotic behavior based on Eq.~\eqref{asy_1} and \eqref{asy_2}. The atom-dimer scattering length $a_{\text{ad}}$ for the $\Lambda$-model is estimated from the trimer binding energy by assuming $E_\text{B}^{\text{(ad)}}\simeq1/2ma_{\text{ad}}^2$. 
	}\label{fig3}
\end{figure}

In Fig.~\ref{fig2}, we display the trimer energy relative to the dimer energy, where we see that the trimer disappears at $a = a^*$ and $+\infty$. 
Of particular interest is the asymptotic behavior of $E_3$ close to the unitarity point. In this limit, the dimer wave function becomes increasingly spatially extended as $1/a\rightarrow0^+$. Thus, 
the repulsion between the dimer state $B_{i\kappa_\text{B}}$ and the additional boson tends to zero, 
and the trimer 
energy in this limit will be $E_3\simeq-2\eb$. Expanding $Z(E_3)=0$ around $-2\eb$, we obtain
\begin{align}
    E_3\simeq\left(-2+\frac{2\pi}{\log a}\right)\eb.\label{asymptotic_3-body}
\end{align}
Thus, 
there is a unique  
logarithmic dependence on scattering length that strongly affects the trimer energy even exponentially close to resonance. This behavior is clearly apparent in Fig.~\ref{fig3}(a). 

To test the universality of the logarithmic term in the trimer energy, we compare our results with those of the ``$\Lambda$-model''.  Here the impurity interacts with bosons via a single-channel contact potential and 
we introduce a momentum cutoff $\Lambda$ in the three-body equation as in Ref.~\cite{Yoshida2018} (see also Ref.~\cite{longpaper}). This is known to be equivalent to including an effective three-body repulsion in the Hamiltonian~\cite{Bedaque1999}. 
As shown in Fig.~\ref{fig3}, we find excellent agreement between these different models, thus demonstrating the universality of our results.

To complete our characterization of the three-body problem, we consider the atom-dimer scattering process in which a boson with incoming momentum $\q$ is scattered by a static dimer. For simplicity, we assume that the total energy $E=\eq-\eb<0$.
In this case, the scattering solution of Eq.~(\ref{3-body eq.}) is~\cite{longpaper}
\begin{align} \notag
    \varphi_{\k\p}=\delta_{\k,\q}\delta_{\p,i\kappa_\text{B}}+\delta_{\k,i\kappa_\text{B}}\delta_{\p,\q}-2\frac{\chi_{\k\p}^*\chi_{\q, i\kappa_\text{B}}/Z(E+i0)}{E-\epsilon_\k-\epsilon_\q+i0}.
\end{align}
where we have shifted $E\to E+i0$ to avoid the divergence and ensure the correct boundary condition at infinity. In the large-distance  
limit, only the atom-dimer part of the wave function, $\varphi_{\rm ad}(\k)\equiv\langle0|b_\k B_{i\kappa_{\rm B}}|\psi\rangle$, remains finite. Using Eq.~(\ref{def_B}) 
and extracting
the large-distance asymptotic behavior,
we obtain an analytic expression for the atom-dimer scattering amplitude~\cite{longpaper}
\begin{align}
    f_{\text{ad}}(\q)=f_0(\q)-e^{2i\delta_q}\frac{\text{Im}[Z(E+i0)]}{q Z(E+i0)}.
    \label{eq:fad}
\end{align}
Here $\delta_q$ is the phase shift for the two channel model, i.e., $q\cot\delta_q=-\frac{1}{a}+\frac{1}{2}r_0q^2$.

From Eq.~\eqref{eq:fad} it is straightforward (although cumbersome) to derive the analytic expressions for low-energy parameters such as 
the atom-dimer scattering length $a_{\text{ad}}$ and the effective range $r_{\text{ad}}$~\cite{longpaper}. In particular, we find that $a_{\text{ad}}$ diverges at the critical scattering length $a^* \simeq |r_0|/0.31821$ 
where the trimer merges with the continuum, as illustrated in Fig.~\ref{fig2}.
We furthermore see that the trimer's binding energy is well approximated by $1/2ma_{\rm ad}^2$,
which indicates that the trimer is nearly universal across its entire region of existence.

\paragraph{$N$-body problem.--}
For larger boson number $N$, it is possible to generalize Eq.~\eqref{finiteU_eq} to an $(N-2)$-fold integral equation for the $(N+1)$-body ground state~\cite{longpaper}.
For instance, in the four-body
problem,
a general state can be written $|\psi\rangle=\sum_{\k\p\q}'\psi_{\k\p\q}B_\k^\dag B_\p^\dag B_\q^\dag|0\rangle$, and the corresponding equation for the tetramer energy $E_4< E_3$ is
\begin{align}
    Z(E_4-\epsilon_\q)f_\q+2{\sum_{\k,\p}}'\frac{\chi_{\k\p}\chi_{\p\q}^*}{E_4-\epsilon_\k-\epsilon_\p-\epsilon_\q}f_\k=0,
\label{eq:tetramer}
\end{align}
where we have defined $f_\q\equiv U\sum_{\u\v}'\chi_{\u\v}\psi_{\u\v\q}$, 
and we have taken the limit $U\to+\infty$.
Figure~\ref{fig2} shows our results for the tetramer energy, 
where we can clearly see that the tetramer ceases to exist at the same scattering lengths as the trimer.

For larger $N$, the calculation of the ground-state energy quickly becomes intractable. 
However, the behavior for scattering lengths 
near unitarity and $a^*$ allows us to infer properties of the many-body system.
By expanding the $(N+1)$-body integral equation around the $a\rightarrow+\infty$ limit, we obtain the asymptotic behavior of the ground-state energy~\cite{longpaper}, 
\begin{align}
    E_{N+1}\simeq\left(-N+\frac{N(N-1)\pi}{\log a}\right)E_\text{B},\quad a\rightarrow +\infty.\label{asy_1}
\end{align}
Thus, for any $N$, we have bound states emerging at the unitary point --- i.e., it is a \emph{multi-body resonance} --- and the ground-state energy features a logarithmic dependence on scattering length similar to that of the trimer energy \eqref{asymptotic_3-body} in this limit. 
As illustrated in Fig.~\ref{fig3}(a), the logarithmic behavior of the multi-body energies is confirmed by our numerical results obtained from solving Eqs.~\eqref{finiteU_eq} and \eqref{eq:tetramer}.
Moreover, we expect Eq.~\eqref{asy_1} to be universal, since we see that it is also captured within the $\Lambda$-model for both trimer and tetramer energies. 
Note that the logarithmic correction grows faster with $N$ than the leading order term, $-N \eb$, thus illustrating the sensitivity to the additional length scale connected to the impurity-induced boson-boson repulsion.

Remarkably, the atom-dimer resonance $a=a^*$ corresponds to a second multi-body resonance. 
Specifically, as $a \to a^*$, the atom-dimer scattering length $a_{\rm ad}$ greatly exceeds the dimer size, 
such that the low-energy physics (measured from $-\eb$) of a system consisting of $N$ bosons and one impurity can be mapped to that of a system consisting of $N-1$ bosons interacting with an infinitely heavy dimer impurity. Thus, the atom-dimer resonance at $a=a^*$
is essentially the unitary point for the boson-dimer system.

Since the $(N+2)$-body problem near $a=a^*$
can be viewed as an $(N+1)$-body problem near the unitary limit, we use 
Eq.~(\ref{asy_1}) to obtain the asymptotic behavior of the energy in the limit $a \rightarrow a^*$,
\begin{align}
    E_{N+2}\simeq-E_\text{B}+\left(-N+\frac{N(N-1)\pi}{\log a_{\text{ad}}}\right)E_\text{B}^{\text{(ad)}},
    \label{asy_2}
\end{align}
where $\eb^\text{(ad)}=|E_3|-\eb$ is the trimer  
binding energy. As seen in Fig.~\ref{fig3}(b), this asymptotic expression well describes the tetramer energy close to $a^*$. Again, this behavior is also found in the $\Lambda$-model, and is thus universal.

The existence of two multi-body resonances implies that bound states with $N >1$ bosons only exist within the interval $0<  1/a  < 1/a^*$, 
i.e., there are no bound states for $a<0$ and there is only the dimer state for $a<a^*$.
Moreover, 
we conjecture that 
all $(N+1)$-body bound states 
exist across the entire interval, 
as depicted in Fig.~\ref{3fig}. 
However, since $\frac{\partial E_{N+1}}{\partial (-1/a)} =  \langle d^\dag d \rangle/ m |r_0| > 0$~\footnote{This is equivalent to the condition that the Tan 
contact~\cite{Tan_2008a} be positive.},
the ground-state energy must decrease monotonically with increasing $|r_0|/a$. Thus, for arbitrary $N$, we have
\begin{align}
    E_{N+1}\geq -\frac{0.038984}{mr_0^2},\quad\text{for }|r_0|/a\leq 0.31821\ldots.
\end{align}
where ${0.038984}/{mr_0^2}$ is the dimer binding energy at $a=a^*$. 
Such an $N$-independent lower bound guarantees that $\lim_{N\rightarrow\infty} |E_{N+2}- E_{N+1}| = 0$, and thus the bound states become increasingly fragile with increasing $N$. We emphasize that this is a universal feature of any impurity system with an effective three-body repulsion.

\paragraph{Implications for the Bose polaron.--}
In the many-body limit $N\rightarrow\infty$, the energy $E_{N+1}$ 
converges to the ground-state polaron energy at zero density (dashed line in Fig.~\ref{3fig}). 
In this case, $a=\infty$ and $a=a^*$ correspond to critical points where the Bose polaron undergoes a sharp transition to/from a ``superpolaronic'' state, where all bosons become bound by the impurity. However, unlike the superpolaronic state in a static scattering potential~\cite{Schmidt2016b}, the polaron is a highly correlated object and its energy is bounded from below.

On the other hand, for a thermodynamic system at finite density, we expect $(N+1)$-body bound states within the multi-body region to be destroyed once $N > N_V$, where $m|E_{N_V+1} - E_{N_V}| \sim (N_V/V)^{2/3}$ with $V$ the system volume.
In other words, the bound state is no longer well-defined once its size becomes larger than the interparticle spacing. Since our results apply to any system with three-body repulsion, such as weak boson-boson repulsion~\cite{Yoshida2018}, we can estimate the typical size of the impurity-induced bound state for the densities in cold-atom experiments. For the weakest interacting BECs, like in Ref.~\cite{Jorgensen2016}, we expect the polaron to bind up to three bosons, while for stronger boson-boson repulsion~\cite{Hu2016}, the polaron can only accommodate two bosons in its dressing cloud. 
This suggests that a heavy impurity in a BEC will behave like a highly correlated object that typically only involves a few bosons.

\begin{acknowledgments}
We gratefully acknowledge fruitful discussions with Xiaoling Cui, Eugene Demler, Shimpei Endo, Victor Gurarie, Hui Zhai.
SMY acknowledges support from  the Japan Society for the Promotion of Science 
through Program for Leading Graduate Schools (ALPS) and 
Grant-in-Aid for JSPS Fellows (KAKENHI Grant No.~JP16J06706).
JL, ZYS, and MMP acknowledge financial support
from the Australian Research Council via Discovery Project
No.~DP160102739. JL is supported through the Australian
Research Council Future Fellowship FT160100244. 
JL and MMP acknowledge funding from the Universities Australia --
Germany Joint Research Co-operation Scheme.
\end{acknowledgments}

\bibliography{bosepolaron}

\end{document}